%% file: paper.tex
\documentclass[twocolumn]{aastex63}
\usepackage{ulem,color}
\usepackage{amsmath,amssymb,bm}
\usepackage{etoolbox}
\usepackage{booktabs}
\usepackage{natbib}
\input{symdef.inc}
\usepackage[normalem]{ulem}

\newcommand{\alf}{Alfv$\acute{\text{e}}$n} %plasma people use Alfven a lot

\usepackage[running]{lineno} 
\AtBeginDocument{%
  \setlength{\linenumbersep}{\dimexpr\oddsidemargin+0.3in-\linenumberwidth\relax}%
}

\received{\today}
\submitjournal{ApJ}
\shorttitle{Waves in inhomogeneous magnetic fields}
\shortauthors{Tripathi \& Mitra}
\graphicspath{{./}{figures/}}
\begin{document}
\title{Exact analytical solutions in inhomogeneous magnetic fields for
  linear asteroseismic waves}
\author[0000-0002-4723-2170]{B. Tripathi}
\affiliation{Nordita, KTH Royal Institute of Technology and Stockholm
  University, Hannes Alfvéns v\"ag 12, 114 19 Stockholm, Sweden}
\affiliation{University of Wisconsin-Madison, Madison, Wisconsin 53706, USA}
\email{btripathi@wisc.edu}

\author[0000-0003-4861-8152]{Dhrubaditya Mitra}
\affiliation{Nordita, KTH Royal Institute of Technology and Stockholm
  University, Hannes Alfvéns v\"ag 12, 114 19 Stockholm, Sweden}
\email{dhruba.mitra@gmail.com}

\begin{abstract}
  We solve for waves in an isothermal, stratified medium with a magnetic field
  that points
  along a direction perpendicular to that of gravity and varies exponentially in the
  direction of gravity.
  We find exact analytical solutions for two different cases: (a) waves
  propagating along the direction
  of the magnetic field and (b) waves propagating along the direction of the gravity.
  In each case, we find solutions in terms of either the hypergeometric functions or their
  confluent cousins.
  We solve the resultant transcendental dispersion relation numerically.
  The changes in the eigenfrequencies due to a spatially inhomogeneous
    \alf\ wave speed are
    significant---there are more oscillations or leaking
      of nodes of the eigenfunctions in regions of lower \alf\ speed and wave
      reflection from regions of larger \alf\ speed.
      Such changes in the dispersion relation and the mode structures may allow
      detection of magnetic fields buried in the
    stellar interior. 

\end{abstract}


\section{Introduction} \label{sec:intro}

The subsurface properties of the Sun have been inferred in great detail by
analyzing the waves on its surface.
This method of helioseismic inversion has revolutionized solar physics by
providing otherwise-inaccessible details of the solar interior, including its
interior rotation profiles (see, e.g., the review by~\citealt{christ2002}).
To create such reconstructions, observational studies commonly use
hydrodynamic properties of the waves.
Although magnetic fields can have additional imprints on the observed surface
oscillations, non-magnetic analyses have offered a great wealth of knowledge
in solar physics~\citep{basu2008, gizon2010}. With
advances in more
precise instruments and calculations, however, observational analysis with
considerations to (evolving) magnetic fields become more important and thus
have received more attention~\citep{jain2007, crouch2005, hanson2014,
  hindman1997, braun1995}.

Theoretical investigations, on the other hand, in global helioseismology,
have been carried out extensively over several decades to predict the effect of magnetic fields on
helioseismic waves, see, e.g., \cite{nye1976, adam1977occurrence,
  thomas1983magneto, campos1983three, lee1986behavior,miles1992magnetoacoustic,
  jain1991magnetoacoustic, cally2007, foullon2005, pinter2011,
  campos2015combined, pinter2015}.
It is well-established that magnetic effects
upshift the frequencies of acoustic
modes of stellar oscillations~\citep{gough1990, goldreich1991,
  dziembowski2004}.
Although these signatures themselves are clear, other studies suggest that these frequency-shifts
may be much smaller than the actually observed range of oscillation
frequencies~\citep{foullon2005}.
This leaves uncertainty in if and how the magnetic fields buried inside the stellar interior can be
reliably estimated.

In local helioseismology, there has been several recent
attempts~\citep{schunker2005, ilonidis2011, singh2014, singh2015pfmodes,
  singh2016high, singh2020} to predict the emergence of active regions via the
detection of large sub-surface magnetic fields through their seismic
signatures. Signatures from numerical simulations~\citep{singh2014,
  singh2015pfmodes, singh2016high, singh2020} and observational studies such as acoustic travel-time
perturbations \citep{schunker2005, ilonidis2011} have been reported to be insightful and in some cases
predictive about the emergence of subsurface magnetic fields on the solar surface and about their
sub-surface field strengths.
In the context of other stars, interior magnetic fields have been assigned bounds, based upon analysis
of depressed dipole oscillation modes due to magnetic fields~\citep{fuller2015asteroseismology}.
Nevertheless, how to decipher the background
profiles of magnetic fields via inversion procedure of astero- and helio-seismic data
remains an unsolved problem.

Waves in magnetized gases, with particular emphasis on solar physics, have been widely studied,
see, e.g.,~\cite{campos1987waves} for a review.
A major bottleneck in progress is the fact that there are
very few exact analytical results for these waves in the presence of  gravity
and an \textit{inhomogeneous magnetic field}.
The first study of MHD waves in an isothermal
      atmosphere~\citep{yu1965} with an inhomogeneous magnetic
      field, assumed a space-dependence such that the \alf\  speed
      remained uniform in space.
      This led to sinusoidal solutions for the linear waves.
      Later work by \cite{nye1976} obtained exact solutions for waves in an
      isothermal atmosphere with a spatially-varying \alf\ wave speed,
      arising from an exponential profile of the mass density and a uniform horizontal magnetic field.
      Even this solution is not completely general---it is
      valid only for waves that propagate along the direction of the magnetic field.
      Note that the inhomogeneity in the magnetic field is undeniably present in
      real stars.
  Roberts and the collaborators~\citep{lee1986behavior,
    miles1992magnetoacoustic, jain1991magnetoacoustic,foullon2005} have
   considered cases where the magnetic field is uniform in one
   layer---in which they use the solution obtained by \cite{nye1976}---and zero in other layer(s) and
   matched the solutions across the boundary
   of the layers. \cite{gough1990} have used perturbation theory to obtain leading-order
   changes to the dispersion relations for the pressure-dominated ($p$-)modes due to 
   non-uniform magnetic fields.
   \cite{singh2015pfmodes} have approached the problem numerically by considering a two-layer
   model---each layer assumed to be isothermal, with different scale heights
   and with a uniform horizontal magnetic field.
   \cite{singh2014} considered the same two-layer isothermal model, but
   with an inhomogeneous magnetic field and found that the inhomogeneity
   significantly changes the amplitude of the fundamental ($f$-) mode of oscillations.
   A recent observational study \citep{korpi2022} suggests that these $f$-mode oscillations may
   have imprints of the poloidal magnetic field as the study exposes anticorrelation of
   mode-integrated energy with the solar cycle, and finds a phase shift of several years.
   Further investigation is needed, though, to ascertain how such coupling of $f$-mode with magnetic
   field occurs.
   Analytically, \cite{campos2015combined} solved for waves
       for a specific profile of magnetic field that decreases exponentially at twice
       the density scale height ($q=2$ where $q$ is the ratio of the two scale heights).
       The exact solutions found were for waves propagating along the direction of gravity.
   The waves propagating along the magnetic field, however, were not analyzed.
   In this article, we consider a magnetic field that varies exponentially with height
   with an arbitrary scale height ($q$ not necessarily equal to two) to find an exact solution for
the waves propagating in two directions---along the
gravity and  along the magnetic field.

Our principal result is: magnetic imprints on the waves, due to inhomogeneous
fields, are quite appreciable, both on the frequencies and the eigenfunctions of
the pressure-dominated modes of oscillations.
The frequencies decrease with increasing degree of the inhomogeneity.
The effects on the eigenfunctions are also identified and node-leakings or
node-shifts are found with spatially-varying magnetic fields.
This causes the eigenfunctions to become broader.
This property may be useful in detection of magnetic fields buried deep
inside a star. 

This article is organized in the following manner.
Section~\ref{sec:sec2} entails the magnetohydrodynamic model of the wave.
A general wave equation is presented in Sec.~\ref{sec:sec3}.
Section~\ref{sec:sec4} contains the details of the background profiles.
Exact solutions for the waves are obtained in Sec.~\ref{sec:sec5}.
In Sec.~\ref{sec:sec6}, effects of an inhomogeneous magnetic
fields on the eigenfrequencies and the eigenfunctions are detailed.
Concluding remarks are made in Sec.~\ref{sec:sec7}.

%------------
\begin{figure}
  \includegraphics[width=0.95\columnwidth]{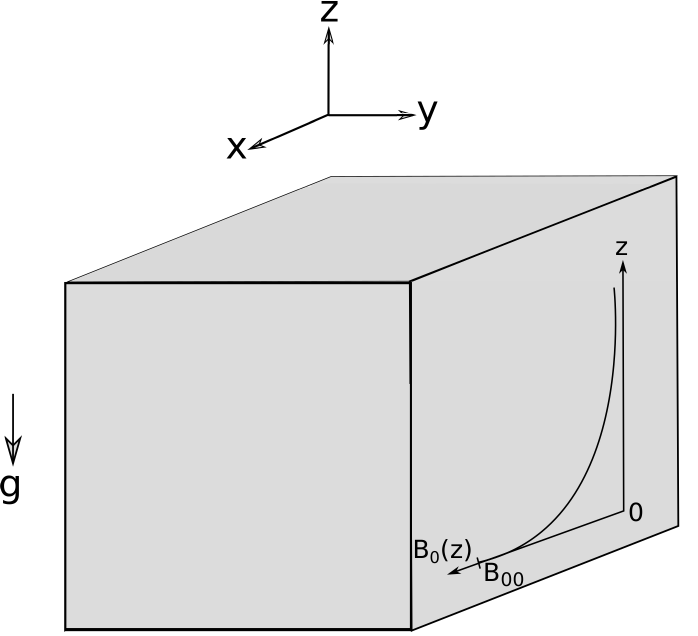} 
  \caption{A schematic diagram of the domain.
    The background velocity is zero and the background magnetic field points along the
    $x$--axis decreasing exponentially with the height $z$.
    The gravity acts vertically downward.
    The horizontal directions $(x,y)$ are periodic.}
\label{fig:sketch}
\end{figure}
% -----------------------------------------------------
\section{Model}\label{sec:sec2}
Consider the equations of motion for magnetized plamas within the framework of
magnetohydrodynamics ~\cite[see, e.g.,][]{choudhuri1998},
with a uniform gravity $\bm{g}$ acting vertically downward:
\begin{subequations}
  \begin{align}
    \delt \rho + \dive \left(\rho \UU \right) &= 0  \/,
    \label{eq:cont} \\
    \rho \left[\delt \UU + (\UU \cdot \bm{\nabla})\UU\right]  &= -\bm{\nabla} P + \rho \bm{g} + \bm{J} \times \bm{B} \/,
    \label{eq:mom}\\
    \delt \BB &= \curl\left( \UU\times\BB \right),\label{eq:induction}
  \end{align}
  \label{eq:MHD}
\end{subequations}
where $\rho$, $\UU$, $P$, and $\BB$ are the density, the velocity, the pressure,
and the magnetic field, respectively.
The current density is $\JJ=\curl\BB/\mun$ with $\mun$ denoting the
permittivity of the vacuum.
Equations~\eqref{eq:cont} and \eqref{eq:induction} are respectively the
continuity equation and the induction equation.
These equations are supplemented with
\begin{equation}
  \dive \BB = 0.
\label{eq:div}
\end{equation}
Dissipative effects such as plasma viscosity and resistivity are
ignored for asteroseismic waves.
These equations must be augmented by an equation of state,
which we consider to be an isothermal condition.
Consequently, the ideal gas law suggests that the sound speed $\cs$, given by
\begin{equation}
  \cs^2 =  \frac{\gmad P}{\rho}\/,
\label{eq:EOS}
\end{equation}
is a constant. The factor $\gamma_{\mathrm{ad}}$ is the adiabatic index of the gas. 

\subsection{Static stationary state}
The plasma with no background flow $\UU=0$ is permeated by a
magnetic field $\BB = \BBn$ that varies in space.
Here, a horizontal magnetic field that decreases vertically with height is
considered, see Fig.~\ref{fig:sketch}.
A similar scenario is relevant in the Sun and Sun-like stars where the
magnetic fields---toroidal in direction (horizontal in local analysis)---decreases
with height.
This vertically varying magnetic field imposes a density profile $\rhon(z)$
where $z$ is the vertical coordinate.
Note a uniform magnetic field or the absence thereof plays no
role in the stationary state.
In such cases, the background density $\rhon(z)$ is a pure exponential
function of $z$, varying with a scale height of $\lrho =\cs^2/(g\gmad) $. 

With a non-uniform magnetic field, the stationary state is given by
\begin{subequations}
  \begin{align}
\label{eq:sMHDrho}\frac{\cs^2}{\gmad} \grad \rhon & = \rho_0 \grav + \JJn\times\BBn
                                   \/,
    \\
    \dive \BBn & = 0 \/, 
    \\
    \JJn &= (\grad \times \BBn)/\mun.
\end{align}
    \label{eq:sMHD}
\end{subequations}
%---------------------------------
\subsection{Linearized MHD equations}
\label{sec:linMHD}
Linear waves are studied by perturbing the governing equations
of motion about a background state as
\begin{equation}
  \left(\rho, \UU, \BB \right) = \left(\rhon, \bm{0}, \BBn \right)
  + \varepsilon\left(\rho, \uu, \bb \right)
  + \mathcal{O}\left(\varepsilon^2\right)\/,
\label{eq:delta}
\end{equation}
where $\varepsilon$ is a small parameter. 

We substitute Eq.~\eqref{eq:delta} in Eq.~\eqref{eq:MHD}, expand it, and keep terms
up to first order in $\varepsilon$.
Algebraic simplifications then reduce the problem to a wave equation for the
perturbed velocity:
\begin{equation}
\begin{aligned} \label{Eperturbation}
\frac{\partial^2 \uu}{\partial t^2} 
= &\frac{1}{\rhon} \nabla (\rhon \cs^2 \nabla \cdot \uu) + \nabla (\uu\cdot \bm{g}) 
- \bm{g}(\nabla \cdot \uu) \\
&-\frac{1}{\mun \rhon} \nabla \left[\uu \cdot \left\{ \nabla \left(\frac{{\Bn}^2}{2}\right) -  \BBn \cdot \nabla \BBn \right\} \right]\\
&+ \frac{1}{\mun \rhon}\left[\nabla \times \left\{ \nabla \times \left(\uu \times \BBn \right) \right\} \right] \times \BBn\\
&+ \frac{1}{\mun \rhon}\left[\left( \nabla \times \BBn \right) \times \left\{ \nabla \times \left(\uu \times \BBn \right) \right\} \right].
\end{aligned}
\end{equation}
This is the standard method of analysis, see, e.g., \cite{chandra1961}.

\subsection{Non-dimensionalization}
\label{sec:ndim}
To simplify the calculations, all the variables henceforth are
non-dimensionalized.
Consider $\lrho$ to be the characteristic length scale,
the speed of sound $\cs$ as the characteristic velocity scale,
the density $\rnn$ and the magnetic field $\Bnn$ at the bottom of the
domain as their respective characteristic scales.
An important dimensionless number $\MaA$, the Alfvenic Mach number, emerges,
see Table~\ref{tab:non}.
Other nondimensional variables are denoted therein by symbols with
asterisks ($\ast$).
In the rest of this article, as we shall use nondimensional units only,
the asterisks are omitted.

%-------------------------
\begin{table}[htbp!]
  \caption {Summary of the non-dimensional parameters. Here, $\lrho$ is the density scale
    height, $\lB$ is the scale height of magnetic-field-variation,
    $\beta$ is the ratio of gas pressure to magnetic
    pressure, and $\MaA$ is the Alfvénic Mach number.}
\hspace{-0.5cm}  
\begin{tabular}{llll}
\toprule
Non-dimensional              &  & Non-dimensional               &  \\
parameter              &  & parameter              &  \\
\midrule
$\zast = \frac{z}{\lrho}$              &  & $\rnast = \frac{\rho_0}{\rnn}$                   &                                                  \\
$D = \lrho \frac{d}{dz}$                      &  & $\Bnast = \frac{{B_0}}{B_{00}}$                        &                                                  \\
$k_x^{\ast} = k_x \lrho$        &  & $\MaA =\frac{c_s}{v_\mathrm{A}}= \frac{\sqrt{\mun \rho_{00}}c_s}{{B_{00}}}$     &  \\
$k_y^{\ast} = k_y \lrho$        &  & $\beta = \frac{P_{\mathrm{gas}}}{P_{\mathrm{mag}}} = \frac{2\mun P_\mathrm{gas}}{{B_{00}}^2}$ &                                      \\
$\omega^{\ast} = \frac{\omega \lrho}{c_s}$ &  &                                                                 $q= \frac{\lB}{\lrho}$ &                                                  \\
\bottomrule
\end{tabular}
\label{tab:non}
\end{table}
%-------------------------
% -----------------------------------------------------
\section{General treatment}\label{sec:sec3}
\label{sec:results}
As the background profiles are homogeneous along the horizontal directions $(x,y)$,
Fourier modes are used along those directions to expand the velocity perturbations
\begin{equation}
  \uu (x,y,z,t) \sim \hat{\uu}(k_x, k_y, z, \omega)\rm{e}^{ i(k_x x+ k_y y - \omega t)},
  \label{eq:uhat}
\end{equation}
where $\hat{\uu}(k_x, k_y, z, \omega)$ is the amplitude of the perturbation at the wavenumbers
$(k_x, k_y)$ with frequency $\omega$.
This expansion when substituted in Eq.~\eqref{Eperturbation} yields a set of three coupled ordinary
differential equations for
the three components of the velocities $\hat{u}_x,\hat{u}_y$, and $\hat{u}_z$.
Using the Einstein convention (summing repeated indices), we find
%---------------------------
\begin{equation} \label{eq:einstein3x3}
    M_{ij} \hat{u}_j + N_{ij} D \hat{u}_j + \delta_{i,3} \left(1 + \frac{{B_0}^2}{\rho_0 \MaA^2} \right) D^2 \hat{u}_3 = 0,
\end{equation}
with $D\equiv d/dz$ and
%-------------------------
%-------------------------
\begin{widetext}
  \begin{eqnarray}
M &=&
\begin{bmatrix}
    \begin{array}{c|c|c}
        \omega^2-{k_x}^2 & -k_x k_y & {-i k_x/ \gmad}  \\ \hline
        -k_x k_y & \omega^2 - {k_y}^2 - k^2  \frac{{B_0}^2}{\rho_0\MaA^2}  & {-i k_y/ \gmad}  \\ \hline
        i k_x \left({\frac{1}{\gmad}-1}-\frac{\gmad}{2\MaA^2}\frac{D{(B_0)}^2}{\rho_0} \right) & i k_y \left({\frac{1}{\gmad}-1}+ \frac{2-\gmad}{2\MaA^2}\frac{D{(B_0)}^2}{\rho_0} \right)  & \omega^2  - \frac{{B_0}^2}{\rho_0\MaA^2}k_x^2
    \end{array}
\end{bmatrix},\\
N &=&
\begin{bmatrix}
    \begin{array}{c|c|c}
        0 & 0 & i k_x  \\ \hline
        0 & 0 & i k_y \left(1 + \frac{{B_0}^2}{\rho_0\MaA^2} \right) \\ \hline
        i k_x & i k_y \left(1 + \frac{{B_0}^2}{\rho_0\MaA^2} \right)  & \frac{2-\gmad}{2\MaA^2} \frac{D{(B_0)}^2}{\rho_0} {- 1}
    \end{array}
\end{bmatrix}.
\end{eqnarray}
\end{widetext}
%----------------

Equation~\eqref{eq:einstein3x3} is general for the case where the background magnetic
field points along the $x$--axis and varies arbitrarily along the $z$-axis.
If it is restricted to be uniform in space, the equations simplify further.
Exact solution of waves for such a profile has already been found~\citet{nye1976,
  adam1977occurrence, thomas1983magneto,campos1983three}.
Here we consider a general case where the background magnetic field
$\BBn$ is an exponential function of the $z$-coordinate.

In particular, we list the following cases:
\begin{enumerate}
 \item \label{case1} waves propagating along the $x$-axis (along the background magnetic field),
 \item \label{case2} waves propagating along the $z$-axis (along the direction of the gravity), and
 \item \label{case3} waves propagating along the $y$-axis (orthogonal to the direction of both the gravity and the background magnetic field).
\end{enumerate}
For waves traveling along the gravity (i.e., case~\ref{case2} above),
\cite{campos2015combined} presented an exact analytical solution for the special case of an
inhomogeneous magnetic field---an exponential function whose scale height is twice
the density scale height, $\BB=\exp(-z/q)\bm{\hat{x}}$ with $q=2$.
In this article, we present exact analytical solutions for both the first and the second case
for magnetic fields that vary exponentially in space \textit{with arbitrary scale height},
i.e., any value of $q$.
  For waves propagating along any arbitrary direction, the problem remains unsolved
  although writing the linearized
  equations themselves is straightforward.
  
  For first two cases outlined above $(k_x, k_y) = (k,0)$, which simplifies
  Eq.~\eqref{eq:einstein3x3} to
\begin{subequations}
\begin{align} 
\begin{split}\label{Ecaseiiia}
{}& \hspace{0cm}\left(\omega^2-{k}^2 \right) \hat{u}_x -i k \left(\frac{1}{\gmad}-D \right)\hat{u}_z =0,
\end{split}\\
\begin{split}\label{Ecaseiiib}
{}&\hspace{0cm} \left(\omega^2 - k^2  \frac{{B_0}^2}{\rho_0\MaA^2} \right)\hat{u}_y = 0,
\end{split}\\
\begin{split}\label{Ecaseiiic}
{}& i k \left[D+\frac{1}{\gmad}-1-\frac{\gmad }{2\MaA^2}\frac{D{(B_0)}^2}{\rho_0} \right] \hat{u}_x\\
&+ \left[\ \left( 1+ \frac{{B_0}^2}{\rho_0\MaA^2}\right) D^2 -\left( 1 + \frac{\gmad-2}{2\MaA^2} \frac{D{(B_0)}^2}{\rho_0}\right) D\right]\hat{u}_z\\
& \hspace{3cm} +\left(\omega^2 - {k}^2 \frac{{B_0}^2}{\rho_0\MaA^2}\right)\hat{u}_z  =0.
\end{split}
\end{align}
\end{subequations}  
Note that $\hat{u}_y$ in Eq.~\eqref{Ecaseiiib} is decoupled from
$\hat{u}_x$ and $\hat{u}_z$. The same equation informs that there exists a mode of oscillation with frequency $\omega^2 = k^2 v_\mathrm{A}^2$ where $v_\mathrm{A}=B_0/(\MaA \sqrt{\rho_0})$. These are, of course, the well-known \alf\ waves. Now, eliminating $\hat{u}_x$ in favor of $\hat{u}_z$ in
Eqs.~\eqref{Ecaseiiia} and \eqref{Ecaseiiic}
\begin{equation}
\begin{aligned} \label{Eworking}
&\left[\left(\omega^2-k^2 \right)\frac{{B_0}^2}{\rho_0\MaA^2} + \omega^2 \right]D^2 \hat{u}_z 
-\omega^2 D\hat{u}_z \\
& \hspace{0cm}+ \left[\left(\omega^2-k^2 \right)\left(\omega^2-k^2  \frac{{B_0}^2}{\rho_0\MaA^2} \right) +\frac{k^2}{\gmad^2} \left(\gmad-1\right) \right] \hat{u}_z\\
&+\frac{D\left({B_0}^2\right)}{\rho_0\MaA^2} \left[\left(\omega^2-k^2-\frac{\omega^2\gmad}{2} \right)D\hat{u}_z+ \frac{k^2}{2}\hat{u}_z \right] = 0.\hspace{0.5cm}
\end{aligned}
\end{equation}
In Eq.~\eqref{Eworking}, the expressions for $\Bn$ and $\rhon$ are not yet prescribed,
which we do next.
If the background magnetic field is uniform, Eq.~\eqref{Eworking} reduces to the one
considered by \cite{nye1976}.
%---------------------------
\subsection{Background profiles} \label{sec:sec4} \label{sec:backgroundprofiles}
Rewriting the force-balance equation for the background state in Eq.~\eqref{eq:sMHDrho}
in non-dimensional form, we obtain
\begin{equation}
D\rho_0 + \rho_0 + \frac{1}{2\MaA^2} \gmad  D(B_0^2)=0\/.
\label{eq:sMHD2}
\end{equation}
Integrating Eq.~\eqref{eq:sMHD2}, we find the background density depends on the background magnetic field via
\begin{equation} 
  \rho_0(z)
  = - \left(\frac{\gmad}{2\MaA^2}\right)\rm{e}^{-z}\int_0^{z} dz'\ \rm{e}^{z'}\ \frac{d(B_0^2)}{dz'} \/.
\label{eq:rnast}
\end{equation}
Equation~\eqref{eq:rnast} is valid for an arbitrary background magnetic field that depends on the
vertical coordinate $z$.
Most previous studies considered a uniform  magnetic field, except, for
instance,~\cite{campos2015combined} who considered a non-uniform magnetic field. They obtained the magnetohydrostatic equilibrium solution
  for a horizontal magnetic field that varies with height. 
  For the solution of waves, they chose a specific profile that decreases
  exponentially at twice the density scale height, $q=2$ in our notation.
  Such a solution is a special case of this study.
In what follows now, the background magnetic field is a generic exponential
function of the height $z$ such that
\begin{equation}
  \label{eq:magprofile}
\BB_0 = \xhat \exp(-z/q),
\end{equation}
where $q = \lB/\lrho$ is the ratio of the characteristic length
scale of magnetic-field-variation to the characteristic length scale of density-variation.
When $q\rightarrow \infty$, the case of the uniform magnetic field is
recovered~\citep{nye1976}.

Substituting Eq.~\eqref{eq:magprofile} in Eq.~\eqref{eq:rnast} and simplifying, we obtain
\begin{subequations}\label{eq:rhoBn}
\begin{align}
\begin{split}\label{eq:rhoqneq2}
{}& \rho_0(z )= \left[1 - \frac{\gmad }{\left(q-2\right) \MaA^2} \right] \rm{e}^{-z} + \frac{\gmad } {\left(q-2\right) \MaA^2} \rm{e}^{-2 z/q}\\ &\mathrm{\hspace{5.2cm} for \hspace{0.2cm}} q \neq 2\/,
\end{split}
\end{align}
\begin{align}
\begin{split}\label{eq:rhoqeq2}
{}&\hspace{-2.12cm} \rho_0 (z) =  \left[1+ \left(\frac{\gmad}{2\MaA^2}\right) z \right]  \rm{e}^{-z} \mathrm{\hspace{0.2cm} for \hspace{0.2cm}} q = 2.\/
\end{split}
\end{align}
\end{subequations}
This means that the density drops exponentially with height for $q \neq 2$.
However, when $q = 2$, it increases linearly with height for $z \ll 1$ and then
drops exponentially for $z \gg 1$.
%------------------
\section{Exact solutions} \label{sec:sec5}
Clearly, the case of $q=2$ is special and turns out to be somewhat simpler.
First, we deal with the case of $q\neq 2$.
\subsection{Alfvén speed for $q\neq 2$ }
\label{sec:case1qn2}
For this case, the density profile is given by Eq.~\eqref{eq:rhoqneq2}.
So, the Alfvén speed, $\Bn^2(z)/\rhon(z)$, can be written as
\begin{equation} \label{EB0squarebyrho0}
v_\mathrm{A}^2 = \left[\frac{\gmad } {\left(q-2\right) \MaA^2} + \left\{1 - \frac{ \gmad }{\left(q-2\right)\MaA^2} \right\} \rm{e}^{-(1-2/q)z} \right]^{-1}\/.
\end{equation}
Straightforward algebra shows
\begin{equation}
\frac{\gmad}{\MaA^2} = \frac{2}{\beta},
\label{eq:beta}
\end{equation}
where $\beta$ is the plasma beta (the ratio of the
gas pressure to the magnetic pressure) at the bottom of the considered domain.
%-------------------
\subsection{Alfvén speed for $q= 2$ }
The Alfvén speed for this case is
\begin{equation}
v_\mathrm{A}^2(z) = \frac{B_0^2}{\rho_0}= {\left[1+ \left(\frac{\gmad}{2\MaA^2}\right) z \right]}^{-1}.
\end{equation}
\subsection{Waves along the magnetic field: $q \neq 2$} 
\label{sec:case1qn3}
Here, we obtain exact solutions for waves that propagate along the magnetic field for
the case of $q \neq 2$.

Substituting Eq.~\eqref{EB0squarebyrho0}
in Eq.~\eqref{Eworking}, simplifying it, and  performing the following change of variable
\begin{equation}
  s = \left(1-\frac{2}{q} \right)z \/,
\label{eq:s}
\end{equation}
equation \eqref{Eworking} assumes the form of
\begin{equation}  \label{EWorking4}
\begin{aligned}
\left[A_2 \rm{e}^{s} + B_2  \right] \frac{d^2\hat{u}_z}{ds^2} &+ \left[A_1 \rm{e}^{s} + B_1 \right] \frac{d \hat{u}_z}{ds}\\
& + \left[A_0  \rm{e}^{s} + B_0  \right] \hat{u}_z = 0,
\end{aligned}
\end{equation}
where the expressions
for $A_0, A_1, A_2$ and $B_0, B_1,B_2$ are given in the Appendix~A.
 
Next, following \cite{campos2015combined}, we perform two more changes of variables,
\begin{subequations}
  \begin{align} \label{eq:xivarchange}
    \xi &= -\frac{B_2}{A_2}\rm{e}^{-s}\/, \\
     W &= \hat{u}_z \rm{e}^{-\theta s}    \/,
  \end{align}
\end{subequations}
where $\theta$ is a yet-undetermined constant.
Upon further simplifications, we obtain
\begin{eqnarray} \label{EWorkingnearfinal}
\begin{aligned}
&\xi (1-\xi) \frac{d^2W}{d\xi^2}\\
& + \left[ \left(2\theta +1 -\frac{A_1}{A_2}\right)-\left(2\theta + 1 -\frac{B_1}{B_2} \right) \xi \right]\frac{dW}{d\xi}\\
&- \left(\theta^2 - \frac{B_1 }{B_2} \theta + \frac{B_0}{B_2} \right) W  + \left(\theta^2 - \frac{A_1}{A_2} \theta + \frac{A_0}{A_2} \right) \frac{W}{\xi} = 0.
\end{aligned}
\end{eqnarray}
The crucial step now is to choose $\theta$ such that the last term in Eq.~\eqref{EWorkingnearfinal}
vanishes, i.e., 
\begin{equation} \label{Ethetacondition}
\theta^2 - \frac{A_1}{A_2} \theta + \frac{A_0}{A_2} = 0 \/.
\end{equation}
Consequently, Eq.~\eqref{EWorkingnearfinal} turns into
the standard hypergeometric differential equation
\begin{equation} \label{EGausshypergeo}
\xi (1-\xi) \frac{d^2W}{d\xi^2} + \left[ C - \left(A + B + 1 \right)
  \xi \right]\frac{dW}{d\xi} - AB W  = 0 \/,
\end{equation}
where the parameters $A$, $B$, and $C$ are given as
 \begin{subequations} \label{EparametersABC}
 \begin{align}
 C &= 2\theta + 1 - \frac{A_1}{A_2},\\
 A + B + 1 &= 2 \theta + 1 - \frac{B_1}{B_2},\\
 AB &= \theta^2 - \frac{B_1 }{B_2} \theta + \frac{B_0}{B_2},
 \end{align}
 \end{subequations}
and $\theta$ is given by Eq.~\eqref{Ethetacondition}.

The solutions to Eq.~\eqref{EGausshypergeo}
are the hypergeometric functions (for $|\xi| < 1 $)
 \begin{equation}
 W\left(A,B;C;\xi \right) = \frac{\Gamma \left(C \right)}{\Gamma \left(A \right) \Gamma \left(B \right)} \sum_{n=0}^{\infty} \frac{\Gamma(A+n) \Gamma(B+n)}{\Gamma(C+n)} \frac{\xi^n}{n!}.
 \end{equation}
Transforming back to the original variables, the final solution is
\begin{equation} 
\begin{aligned}\label{Efinaleqn}
&\hat{u}_z(z)\\
&= D_1 \ \rm{e}^{-z \theta  (1-2/q)} \  {}_{2}{F}_{1}\left(A,B;C;-\frac{B_2}{A_2} \rm{e}^{-z (1-2/q)}\right)\\
&+ D_2 \  \rm{e}^{z \theta (1-2/q) (1-\theta A_1/A_2 )} {\left[\frac{-B_2}{A_2}\right]}^{1-C}\\
&\times {}_2 F_1\left(A-C+1, B-C+1; 2-C; -\frac{B_2}{A_2} \rm{e}^{-z (1-2/q)}\right)\/,
\end{aligned}
\end{equation}
where $D_1$ and $D_2$ are constants. They are constrained by the boundary conditions. The function ${}_{2}{F}_{1}$ is the Gauss hypergeometric function.
All constants $A$, $B$, $C$, $A_0$, $A_1$, $A_2$, $B_0$, $B_1$, $B_2$, and
$\theta$ are functions of $\omega$, $k$, $\gmad$, $\MaA$, and $q$
(which are given in the Appendix~A).
%--------------------------
\subsection{Waves along the magnetic field: $q = 2$}
\label{sec:case1qeq2}
Following a similar procedure as in Sec.~\ref{sec:case1qn3}, we obtain
\begin{equation} \label{Efinalsolutiontwice}
 \begin{aligned}
  \hat{u}_z(k, z, \omega) &= A_{\Plus}\  {}_{1}F_{1}\left(m_{\Plus};1; \Minus \eta_4 \left(\eta_1 + z \right) \right) \ \rm{e}^{\left( \frac{1\Plus \eta_4}{2} \right) \left(\eta_1 + z \right)}\\
  & + A_{\Minus}\  {}_{1}F_{1}\left(m_{\Minus};1;  \eta_4 \left(\eta_1 + z \right) \right) \ \rm{e}^{\left( \frac{1\Minus \eta_4}{2} \right) \left(\eta_1 + z \right)},
  \end{aligned}
 \end{equation}
where the function ${}_{1}F_{1}$ is the confluent hypergeometric
function of the first kind. The relation between the constants $A_\Plus$ and $A_\Minus$ are determined by the choice of the boundary conditions.
The constants $m_{\Plus}$ and $m_{\Minus}$ are
\begin{eqnarray} \label{Epsandqs}
 \begin{aligned}
 m_\pm &= \frac{1}{2}\left(1 \pm\frac{1}{\eta_4} \right) \pm \frac{\eta_2}{\eta_4} \left( \eta_3-\eta_1\right).
%  n &= 1.
 \end{aligned}
 \end{eqnarray}
The functions represented by $\eta_j$ with $j=1,2,3,4$ are functions of $\omega$, $k$, $\gmad$, and $\beta$ as given in the Appendix~B.
\subsection{Waves along the magnetic field:
  $q \rightarrow \infty$ (uniform $\textbf{B}_0$)}

If $\BBn$ is uniform in space, we recover the solution of \cite{nye1976}
from our solution by taking the limit $q\to \infty$ in Eq.~\eqref{Efinaleqn}.
In this limit, the parameters simplify to
\begin{eqnarray} \label{eq:paramnye}
\begin{aligned}
A_2 &= a_2 = (\omega^2 - k^2) /\MaA^2,\\
A_1 &= a_1 = 0,\\
A_0 &= a_0 = -k^2 (\omega^2 -k^2)/\MaA^2.
\end{aligned}
\end{eqnarray}
Equation~\eqref{Ethetacondition} also reduces to
 \begin{equation}
 \theta =  k\/.
\end{equation}
   This set of parameters in Eq.~\eqref{eq:paramnye}, when substituted in
   Eq.~\eqref{Efinaleqn}, reproduces the solution obtained by~\cite{nye1976}.

\subsection{Waves propagating along the direction of the gravity} \label{sec: wave_gravity}
Next we consider the problem of waves propagating perpendicular to the magnetic field and
along the direction of the gravity.
We separate the two cases: $q\neq 2$ and $q=2$.
The latter has already been solved exactly by \cite{campos2015combined}.
Below we present the solution for the former.

As the waves propagate along the $z$-axis, $k_x=k_y=0$. This when substituted in
Eq.~\eqref{eq:einstein3x3} yields
\begin{subequations}
\begin{align}
\begin{split}
\hspace{3.6cm}\omega^2 \hat{u}_x =0&,  \\
\end{split}
\end{align}
\begin{align}
\begin{split}
\hspace{3.6cm} \omega^2 \hat{u}_y = 0&, \\
\end{split}
\end{align}
\begin{align}
\begin{split}
\label{eq: verticalmag3} \left[\omega^2 + D^2\left(1 + \frac{{B_0}^2}{\rho_0\MaA^2}\right) - D + \frac{2-\gmad}{2\MaA^2} \frac{D{(B_0)}^2}{\rho_0} D \right]&\hat{u}_z\\
 &\hspace{-2.1cm}=0 \/.
\end{split}
\end{align}
\end{subequations}
The solution to Eq.~\eqref{eq: verticalmag3}
can be found as a special case by substituting $k_x = k = 0$ in the solution presented in
Sec.~\ref{sec:case1qn3}.
The solution is
\begin{equation}  \label{eq: verticalmagsoln}
\begin{aligned}
&\hat{u}_z(z)\\
&= D_1 \ \rm{e}^{-z \theta  (1-2/q)} \ {}_{2}{F}_{1} \left(A,B;C;-\frac{B_2}{A_2} \rm{e}^{-z (1-2/q)}\right)\\
&+ D_2 \  \rm{e}^{z \theta (1-2/q) (1-\theta A_1/A_2)} {\left[\frac{-B_2}{A_2}\right]}^{1-C}\\
& \times {}_2 F_1\left(A-C+1, B-C+1; 2-C; -\frac{B_2}{A_2} \rm{e}^{-z (1-2/q)}\right),
\end{aligned}
\end{equation}
where all the arguments of the hypergeometric functions are redefined in the Appendix~C.

The special solution for $q = 2$, presented by
\cite{campos2015combined}, can be obtained by substituting
$k = 0$ in Eq.~\eqref{Efinalsolutiontwice}.
%-----------------
\section{Effect of an inhomogeneous magnetic fields on waves}\label{sec:sec6}
\begin{figure}[htbp!]
	\includegraphics[width=0.47\textwidth]{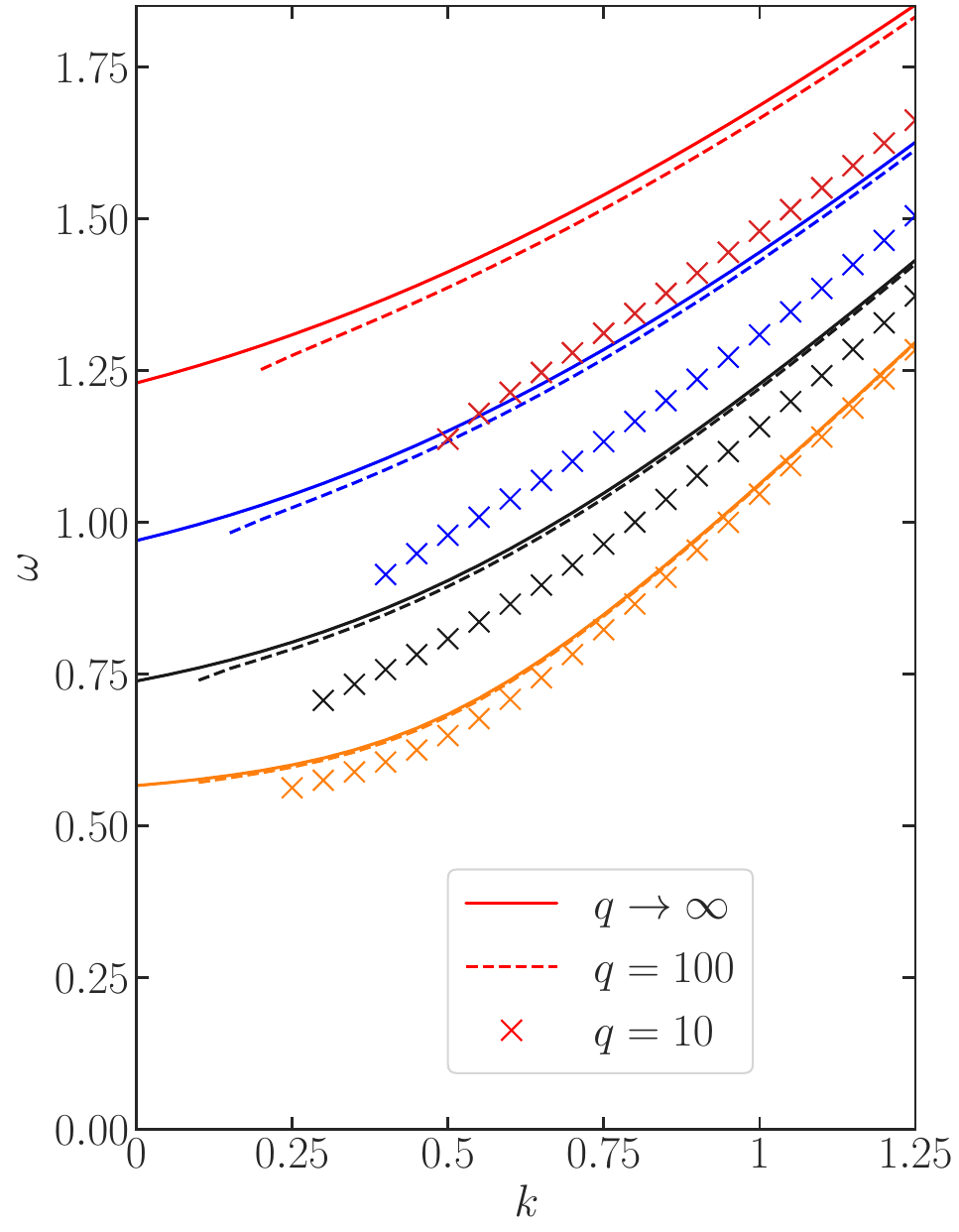}
        \caption{\label{fig:fig2} Dispersion relation for
          waves in a magnetized plasma with a horizontal magnetic field
          that decreases exponentially with the height.
          Here, $q$ is the scale height of
          magnetic-field-variation over 
          the scale height of density-variation.
          The case $q \rightarrow \infty$ corresponds to a uniform magnetic
          field. We choose
          $\gamma_{\mathrm{ad}}=5/3$ and $\MaA^{-2}=10^{-4}$ in order to compare
          with the dispersion relation of \cite{nye1976}.  
We also use $q=100$ and $q=10$. First four $p$--modes are shown.
The rapid decrease in the magnetic field with
the height significantly lowers the wave frequencies,
particularly those of the higher oscillation modes.
}
\end{figure}
\begin{figure*}[h]
	\includegraphics[width=0.96\textwidth]{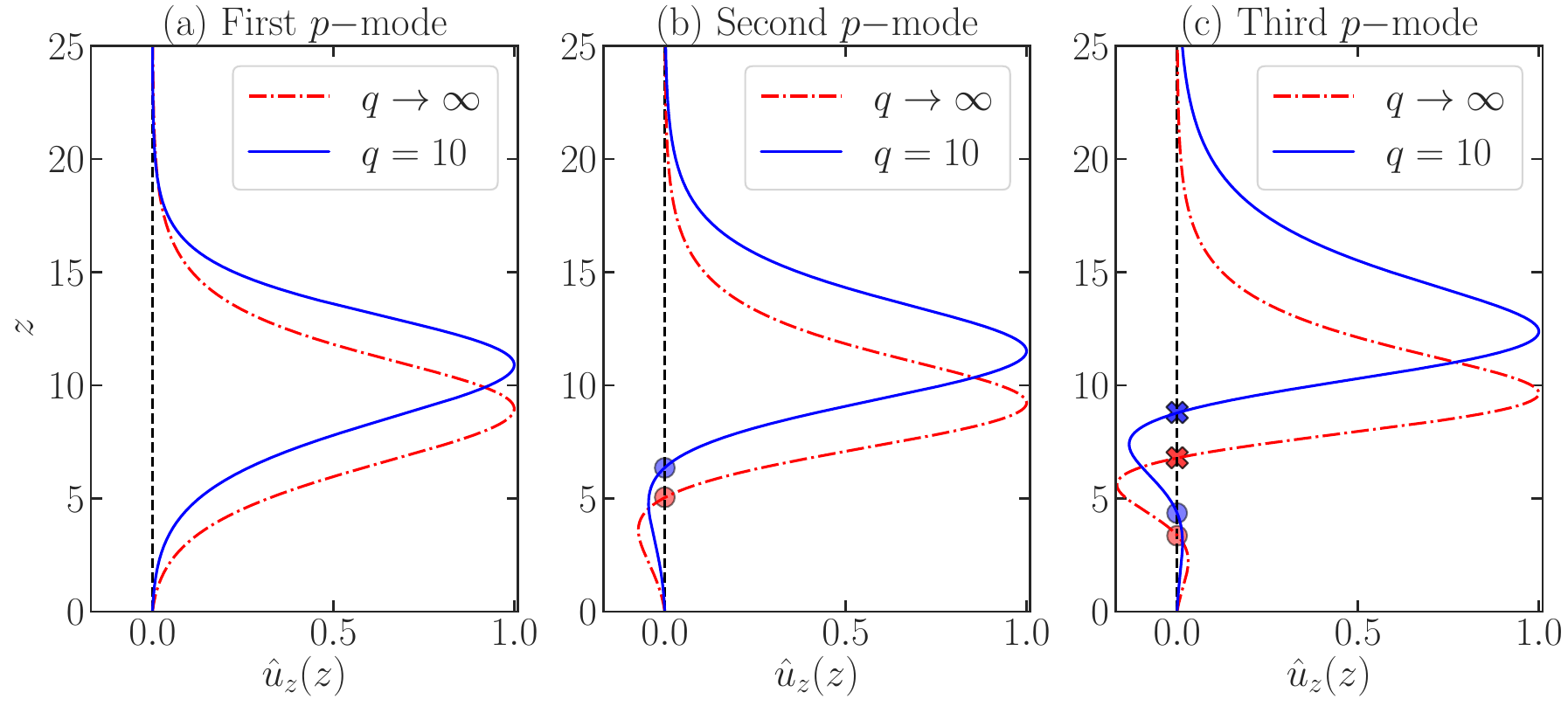}
        \caption{Leakage of eigenfunctions (or nodes) towards the region of weaker magnetic
          fields (i.e., higher $z$ where the \alf\ speed is reduced with smaller values of $q$).
          The parameter $q=10$ is the scale height of the magnetic field in units of the density scale
          height.
          The case $q\rightarrow \infty$ corresponds to a uniformly magnetized plasma.
          Shown eigenfunctions are for (a) the first, (b) the second, and (c) the third modes of
          pressure-dominated oscillations at wavenumber $k=0.5$.
          Their respective eigenfrequencies are $\omega = 0.6835, 0.9037, 1.1500
          $ (for $q\rightarrow \infty$) and $\omega = 0.6484, 0.8079, 0.9784 $ (for $q=10$).
          Other parameters used are $\gamma_{\mathrm{ad}}=5/3$ and $\MaA=100$.
          Note that the nodes of the eigenfunctions penetrate deeper (i.e., larger $z$-values)
              in the region where the \alf\ speed is reduced as $q$ is changed from
               $q \rightarrow \infty$ to $q=10$. }
    \label{fig:fig3}
\end{figure*}
%--------------------------
Here, we obtain the dispersion relations of waves in non-uniformly magnetized plasmas and
compare them with well-known waves in uniformly magnetized
plasmas.
%Since the magnetic fields considered here decay with height, waves \BT{\sout{in upper regions
%can ``sense" the stronger magnetic fields lying below (or afar).
%How sensitive they are is investigated next} waves may ``sense" the strength of the ambient magnetic field (or the \alf\ wave speed) and penetrate in regions of larger \alf\ wave speed correspondingly. This, in other words, is a problem of wave reflection and penetration. How sensitive these waves are to the inhomogeneous magnetic field is of interest to investigate next}.
By imposing the boundary conditions on the analytical solution found in
Eq.~\eqref{Efinaleqn}, the dispersion relations are derived.
Following \cite{nye1976}, we also demand that the energy per--unit-volume of the perturbed fields to be
remain bounded. 
This implies $D_2=0$ in Eq.~\eqref{Efinaleqn}.
Note that the upper boundary at $z\rightarrow \infty$ maps to $\xi=0$ in the new
coordinate $\xi$, defined in Eq.~\eqref{eq:xivarchange}.
Imposing a rigid lower boundary at $z=0$ ($\xi=-B_2/A_2$), where the vertical
velocity $\hat{u}_z$ vanishes, the dispersion relation is given by
\begin{equation}
    {}_{2}{F}_{1} \left(A(\omega,k,q),B(\omega,k,q);C(\omega,k,q);-\frac{B_2(\omega,k,q)}{A_2(\omega,k,q)}\right)=0,
\end{equation}
where all the parameters $A, B, C, A_2, B_2$ are functions of $\omega$, $k$, and $q$.
Their expressions are given in Eq.~\eqref{EparametersABC} and in the Appendix~A. 

In Fig.~\ref{fig:fig2}, the dispersion relations for the waves in uniformly and
non-uniformly magnetized plasmas are compared.
Decreasing magnetic field with the increasing height has the effect of lowering the
oscillation frequencies of the waves---which is particularly pronounced for the higher oscillation modes. 

Now, we show the effect of spatially-varying magnetic fields on the eigenfunctions of
the oscillations.
As seen in Fig.~\ref{fig:fig3},
the eigenfunctions leak towards (penetrate into) the upper regions,
i.e., higher $z$-values, where the magnetic fields are weaker.
For example, the second and the third modes of oscillations show that their nodes are shifted
due to weakening of the magnetic fields.
This may alternatively be interpreted as lowering of the \alf\ speed in these regions
    when the magnetic field is non-uniform ($q=10$), compared to when it is
    uniform ($q \rightarrow \infty$).
    This lowered \alf\ speed leads to a softened (lowered) wave reflection,
    thus enhancing oscillations in these regions \citep{campos2015combined}.

In this article, we have considered the background inhomogeneous magnetic
field to be constant-in-time.
This is physically justifiable as the time scale of emergence of
magnetic flux and the formation of the active
regions on the solar surface is very slow compared to the characteristic
time scales of the helioseismic waves. Hence it is a good enough approximation
to consider the background magnetic field to be static.
Typical emergence phenomenon is the formation of a bipolar region,
which is expected to form from the underlying magnetic fields that
are horizontal.
Thus starting with a horizontal magnetic field (to represent the fields lying
deep inside) can also be taken as a reasonable approximation.
But the assumption of isothermal atsmosphere does not allow us
to compare our results directly with solar or stellar observations~\citep{ulrich1970five}. 

% -----------------------------------------------------
\section{Conclusions}\label{sec:sec7}
We have presented hitherto-unknown exact analytical solutions for the linear
asteroseismic waves, propagating through a medium with a
gravity and an exponentially-varying horizontal magnetic field, whose scale height is arbitrary.
These solutions and the wave dispersion relations
are found to be sensitive to the degree of the
inhomogeneity.
To whit, the eigenfrequencies of the waves rapidly decrease when the magnetic field strength decreases
faster with height.
The eigenfunctions become broader, with their nodes leaking towards the
regions of weaker magnetic fields.
This is understood as softened wave reflection from a region where the
    \alf\ wave speed gets lowered.

It may be noted that mathematically similar problem as considered here appears while analyzing the generation of ocean waves by a turbulent wind of exponential profile~\citep{miles1960generation}.
Recently, perturbative solutions for the Miles problem and also
for the case of a wind with other profiles (in particular, logarithmic)
were obtained using the method of matched asymptotics~\citep{bonfils2021asymptotic}.
A variant of such a technique may allow to obtain perturbative solutions for more complicated profile of inhomogeneous background magnetic field. But, the insight about the interplay of waves, their nodes, and the inhomogeneous \alf\ speed may carry over from the problem discussed in this article.

Apart from informing the observational studies, the solution found here for the
linear eigenmodes can be employed to compute nonlinear mode coupling coefficients that can test and
verify numerical codes of nonlinear asteroseismology.
In other words, having an exact analytical solution with non-trivial magnetic field
geometry helps to set stringent constraint on
benchmarking numerical codes of
helio- or astero-seismology.
Further, asymptotics of these exact solutions can offer insights into
the imprints of nonlinearities on the observed stellar oscillations.
Such an investigation will be the focus of a future study.

\section*{Acknowledgments}
We thank E.G.~Zweibel for reading through the manuscript and for providing helpful comments. B.T. takes pleasure in thanking J.~Fuller and other participants of the program, ``Transport in Stellar Interiors, 2021," at the Kavli Institute of Theoretical Physics for useful discussions. Thanks also to the anonymous reviewer for insightful suggestions that helped to improve the conclusions of the manuscript. For offering a Visiting Fellowship, B.T. is indebted to the Nordic Institute for Theoretical Physics, Sweden where most of the work was performed. D.M. is supported by the Swedish Research Council Grant No.~638-2013-9243. D.M. gratefully acknowledges P.~Kumar for help in initial stages of this work.
%____

%\section*{Appendix A: Expressions for $C_0$, $C_1$, $C_2$, and $C_3$}
%\ATT{Rewrite the following using the factor 4 times gamma squared in equation \eqref{eq: omegasconstB}. We present here the expressions for terms ($C$'s) employed in equation \eqref{eq: omegasconstB} of section \ref{sec:constntBdisp}.
%\begin{subequations} 
%\begin{align*}
%\begin{split}
%{}& C_0 = - 4 {\omega_M}^2 \left[{\omega_M}^2 \left({\omega_a}^2 -3 \gamma^2/4 -1/4 \right) + {\omega_I}^2 {\omega_a}^2 \right]
%\end{split}\\
%\begin{split}
%{}& C_1 = \left( 1 + \frac{{\omega_M}^2}{{\omega_a}^2 - \gamma^2 - 1/4} \right) \left\{ 4 {\omega_M}^2 \left({\omega_a}^2 - \gamma^2/2 - 1/4 \right) + 4 {\omega_I}^2 {\omega_a}^2  \right\}\\
%& \hspace{2cm} + 4 {\omega_M}^2 \left({\omega_M}^2 + {\omega_a}^2 - \gamma^2 + 1/4 \right)
%\end{split}\\
%\begin{split}
%{}& C_2 = - 4\left( {\omega_M}^2 + {\omega_a}^2 - \gamma^2 - 1/4 \right) \left( 1 + \frac{{\omega_M}^2}{{\omega_a}^2 - \gamma^2 - 1/4} \right)\\
%& \hspace{0.5cm} - 4 {\omega_M}^2 - \frac{{2 \omega_M}^2}{{\omega_a}^2 - \gamma^2 - 1/4} - 1 - \gamma^2 {\left(\frac{{\omega_M}^2}{{\omega_a}^2 - \gamma^2 - 1/4} \right)}^2
%\end{split} \\
%\begin{split}
%{}& C_3 = 4 \left( 1 + \frac{{\omega_M}^2}{{\omega_a}^2 - \gamma^2 - 1/4} \right)
%\end{split}
%\end{align*}
%\end{subequations}}

\section*{Appendix A: Expressions for $A_{\MakeLowercase{j}}$ and $B_{\MakeLowercase{j}}$} \label{sec:appendixA}
The expressions for the parameters introduced in Eq.~\eqref{EWorking4} of Sec.~\ref{sec:case1qn3} are presented below:
\begin{eqnarray*} 
\begin{aligned}
A_0 &= \frac{1}{\MaA^2}\left[\left(\omega^2-k^2\right)\left(\frac{\omega^2\gmad} {q-2} -k^2  \right) + k^2 \frac{\left(\gmad-1\right)}{\gmad \left(q-2\right)} - \frac{k^2}{q}\right],\\
A_1 &= \frac{1}{\MaA^2}\left[ \frac{2}{q}\left(-\omega^2+k^2+\frac{\omega^2\gmad}{2} \right)+ \omega^2 \frac{\gmad} {\left(q-2\right)} \right]{\left(1-\frac{2}{q} \right)}, \\
A_2 &= \frac{1}{\MaA^2}\left[\omega^2-k^2 +  \frac{\omega^2\gmad} {\left(q-2\right)}\right] {\left(1-\frac{2}{q} \right)}^2,\\
B_0 &= \ \ \omega^2  \left[1 - \frac{ \gmad }{\left(q-2\right)\MaA^2} \right] \left[ \omega^2-k^2   + \frac{k^2}{\omega^2} \frac{\left(\gmad-1\right)}{\gmad^2}\right],\\
B_1 &= - \omega^2 \left[1 - \frac{ \gmad }{\left(q-2\right)\MaA^2} \right] {\left(1-\frac{2}{q} \right)},\\
B_2 &= \ \ \omega^2 \left[1 - \frac{ \gmad }{\left(q-2\right)\MaA^2} \right]{\left(1-\frac{2}{q} \right)}^2.
\end{aligned}
\end{eqnarray*}

\section*{Appendix B: Expressions for $\eta_{\MakeLowercase{j}}$}
The expressions for $\eta_j$ used in Eq.~\eqref{Epsandqs} of Sec.~\ref{sec:case1qeq2} are listed here:
\begin{eqnarray*}
\begin{aligned}
\eta_1 &= \frac{2}{\gmad} \left(1-\frac{k^2 }{\omega^2}\right)  + \beta,  \\ % \eta_2 &= 1-\eta_1 \\
\eta_2 &=  \frac{k^2}{\omega^2} \frac{\left(\gmad-1\right)}{\gmad^2}  + \left(\omega^2-k^2 \right), \\
\eta_3 &= \frac{\left(1-\frac{k^2}{\omega^2} \right) \left(\omega^2\beta -\frac{2k^2}{\gmad} \right) +\frac{k^2}{\omega^2} \frac{\left(\gmad-1\right)}{\gmad^2} \beta - \frac{k^2}{\omega^2}\gmad }{\eta_2}, \\
\eta_4 &= {\left(1-4\eta_2\right)}^{1/2}.
\end{aligned}
\end{eqnarray*}

\section*{Appendix C: Expressions for $A_{\MakeLowercase{j}}$, $B_{\MakeLowercase{j}}$, and $C_{\MakeLowercase{j}}$}
The parameters that appear in Eq.~\eqref{eq: verticalmagsoln} of Sec.~\ref{sec: wave_gravity} are given next:
\begin{eqnarray*}
\begin{aligned}
A_0 &= \frac{1}{\MaA^2} \frac{\omega^4\gmad} {q-2},  \\
A_1 &= \frac{1}{\MaA^2}\omega^2 \left(\frac{\gmad-2}{q} -\frac{\gmad } {q-2}  \right){\left(1-\frac{2}{q} \right)},\\
A_2 &= \frac{1}{\MaA^2} \omega^2\left( 1  +  \frac{\gmad} {q-2}\right){\left(1-\frac{2}{q} \right)}^2,\\
B_0 &=  \ \ \omega^4 \left[1 - \frac{ \gmad}{\left(q-2\right)\MaA^2} \right],\\
B_1 &= - \omega^2 \left[1 - \frac{ \gmad}{\left(q-2\right)\MaA^2} \right]{\left(1-\frac{2}{q} \right)},\\
B_2 &= \ \ \omega^2 \left[1 - \frac{ \gmad }{\left(q-2\right)\MaA^2} \right]{\left(1-\frac{2}{q} \right)}^2.  
\end{aligned}
\end{eqnarray*}

Another parameter $\theta$ is evaluated using the Eq.~\eqref{Ethetacondition}, which yields
\begin{equation*} 
\theta = \frac{1}{2}\left(\frac{A_1}{A_2} \pm \sqrt{{\left(\frac{A_1}{A_2}\right)}^2 -\frac{4A_0}{A_2}}\right).
\end{equation*}

Lastly, the parameters $A$, $B$, and $C$ in Eq.~\eqref{eq: verticalmagsoln} are found by solving the following set of equations
 \begin{eqnarray*} 
 \begin{aligned}
 C &= 2\theta + 1 - \frac{A_1}{A_2},\\
 A + B + 1 &= 2 \theta + 1 - \frac{B_1}{B_2},\\
 AB &= \theta^2 - \frac{B_1 }{B_2} \theta + \frac{B_0}{B_2}.
 \end{aligned}
 \end{eqnarray*}
This completes the specification for obtaining the Gauss hypergeometric function as the solution for the magneto-acoustic-gravity waves in a medium permeated by spatially-varying magnetic fields.

\vspace{6cm}

%% This command is needed to show the entire author+affiliation list when
%% the collaboration and author truncation commands are used.  It has to
%% go at the end of the manuscript.
%\allauthors

%% Include this line if you are using the \added, \replaced, \deleted
%% commands to see a summary list of all changes at the end of the article.
%\listofchanges

% %--------------------------
% \bibliography{turb_ref,sunref}
% \bibliographystyle{mn2e}

% \input{paper.bbl} 
% \bibliography{sunref}
% \bibliographystyle{aasjournal}

%--------------------------

\bibliography{references_aastex}{}
\bibliographystyle{aasjournal}

\end{document}

%% file: symdef.inc
\def \Bnast {B_{\rm 0}^{\ast}}
\def \rnast {\rho_{\rm 0}^{\ast}}
\def \rnn  {\rho_{\rm 00}}
\def \Bnn  {B_{\rm 00}}
\def \zast {z^{\ast}}
\def \Bn  {B_{\rm 0}}
\def \BBn {\bm{B}_{\rm 0}}
\def \JJn {\bm{J}_{\rm 0}}
\def \lrho {\ell_{\rho}}
\def \lB   {\ell_{\rm B}}
\def \rhon {\rho_{\rm 0}}

\def \xhat {\hat{x}}

\def \gmad {\gamma_{\rm ad}}

\def \del2z {\partial^{2}_{z}}

\def \A {{\bm A}}

\def \f {{\bm f}}

\def \uu {{\bm u}}
\def \bb {{\bm b}}
\def \UU {{\bm U}}

\def \B {{\bm B}}

\def \JJ {{\bm J}}
\def \mun {\mu_{{\rm 0}}}

\def \k {{\bm k}}
\def \x {{\bm x}}
\def \grav {{\bm g}}
\def \grad {{\bm \nabla}}
\def \curl {{\bm \nabla} \times}
\def \dive {{\bm \nabla}\cdot}

\def \delt {\partial_t}

\def \t  {\theta}
\def \p  {\phi}

{}

\def \MaA  {\mbox{M}_{\rm A}}

\def \BB {\bm B}
\def \B {\bm B}

\def \A {\bm A}

\def \curl {{\bm \nabla}\times}

\def \cs  {c_{\rm s}}

\def \BB {\bm B}

 \def \a {{\alpha_0}}

\def \Plus{\texttt{+}}
\def \Minus{\texttt{-}}
\def\drawing #1 #2 #3 {
\begin{center}
\setlength{\unitlength}{1mm}
\begin{picture}(#1,#2)(0,0)
\put(0,0){\framebox(#1,#2){#3}}
\end{picture}
\end{center} }

%% file: paper.bbl
\begin{thebibliography}{}
\expandafter\ifx\csname natexlab\endcsname\relax\def\natexlab#1{#1}\fi
\providecommand{\url}[1]{\href{#1}{#1}}
\providecommand{\dodoi}[1]{doi:~\href{http://doi.org/#1}{\nolinkurl{#1}}}
\providecommand{\doeprint}[1]{\href{http://ascl.net/#1}{\nolinkurl{http://ascl.net/#1}}}
\providecommand{\doarXiv}[1]{\href{https://arxiv.org/abs/#1}{\nolinkurl{https://arxiv.org/abs/#1}}}

\bibitem[{Adam(1977)}]{adam1977occurrence}
Adam, J. 1977, Solar Physics, 52, 293

\bibitem[{{Basu} \& {Antia}(2008)}]{basu2008}
{Basu}, S., \& {Antia}, H.~M. 2008, \physrep, 457, 217,
  \dodoi{10.1016/j.physrep.2007.12.002}

\bibitem[{Bonfils {et~al.}(2021)Bonfils, Mitra, Moon, \&
  Wettlaufer}]{bonfils2021asymptotic}
Bonfils, A., Mitra, D., Moon, W., \& Wettlaufer, J. 2021, arXiv preprint
  arXiv:2107.06844

\bibitem[{{Braun}(1995)}]{braun1995}
{Braun}, D.~C. 1995, \apj, 451, 859, \dodoi{10.1086/176272}

\bibitem[{{Cally}(2007)}]{cally2007}
{Cally}, P.~S. 2007, Astronomische Nachrichten, 328, 286,
  \dodoi{10.1002/asna.200610731}

\bibitem[{Campos(1983)}]{campos1983three}
Campos, L. 1983, Wave Motion, 5, 1

\bibitem[{Campos(1987)}]{campos1987waves}
---. 1987, Reviews of Modern Physics, 59, 363

\bibitem[{Campos \& Marta(2015)}]{campos2015combined}
Campos, L., \& Marta, A. 2015, Geophysical \& Astrophysical Fluid Dynamics,
  109, 168

\bibitem[{Chandrasekhar(1961)}]{chandra1961}
Chandrasekhar, S. 1961, Hydrodynamic and hydromagnetic stability, 652 pp.,
  clarendon,  Oxford, UK

\bibitem[{Choudhuri {et~al.}(1998)}]{choudhuri1998}
Choudhuri, A.~R., {et~al.} 1998, The physics of fluids and plasmas: an
  introduction for astrophysicists (Cambridge University Press)

\bibitem[{{Christensen-Dalsgaard}(2002)}]{christ2002}
{Christensen-Dalsgaard}, J. 2002, Reviews of Modern Physics, 74, 1073,
  \dodoi{10.1103/RevModPhys.74.1073}

\bibitem[{{Crouch} \& {Cally}(2005)}]{crouch2005}
{Crouch}, A.~D., \& {Cally}, P.~S. 2005, \solphys, 227, 1,
  \dodoi{10.1007/s11207-005-8188-z}

\bibitem[{{Dziembowski} \& {Goode}(2004)}]{dziembowski2004}
{Dziembowski}, W.~A., \& {Goode}, P.~R. 2004, \apj, 600, 464,
  \dodoi{10.1086/379708}

\bibitem[{{Foullon} \& {Roberts}(2005)}]{foullon2005}
{Foullon}, C., \& {Roberts}, B. 2005, \aap, 439, 713,
  \dodoi{10.1051/0004-6361:20041910}

\bibitem[{Fuller {et~al.}(2015)Fuller, Cantiello, Stello, Garcia, \&
  Bildsten}]{fuller2015asteroseismology}
Fuller, J., Cantiello, M., Stello, D., Garcia, R.~A., \& Bildsten, L. 2015,
  Science, 350, 423

\bibitem[{{Gizon} {et~al.}(2010){Gizon}, {Birch}, \& {Spruit}}]{gizon2010}
{Gizon}, L., {Birch}, A.~C., \& {Spruit}, H.~C. 2010, \araa, 48, 289,
  \dodoi{10.1146/annurev-astro-082708-101722}

\bibitem[{{Goldreich} {et~al.}(1991){Goldreich}, {Murray}, {Willette}, \&
  {Kumar}}]{goldreich1991}
{Goldreich}, P., {Murray}, N., {Willette}, G., \& {Kumar}, P. 1991, \apj, 370,
  752, \dodoi{10.1086/169858}

\bibitem[{Gough \& Thompson(1990)}]{gough1990}
Gough, D., \& Thompson, M. 1990, Monthly Notices of the Royal Astronomical
  Society, 242, 25

\bibitem[{{Hanson} \& {Cally}(2014)}]{hanson2014}
{Hanson}, C.~S., \& {Cally}, P.~S. 2014, \apj, 791, 129,
  \dodoi{10.1088/0004-637X/791/2/129}

\bibitem[{{Hindman} {et~al.}(1997){Hindman}, {Jain}, \&
  {Zweibel}}]{hindman1997}
{Hindman}, B.~W., {Jain}, R., \& {Zweibel}, E.~G. 1997, \apj, 476, 392,
  \dodoi{10.1086/303615}

\bibitem[{{Ilonidis} {et~al.}(2011){Ilonidis}, {Zhao}, \&
  {Kosovichev}}]{ilonidis2011}
{Ilonidis}, S., {Zhao}, J., \& {Kosovichev}, A. 2011, Science, 333, 993,
  \dodoi{10.1126/science.1206253}

\bibitem[{{Jain}(2007)}]{jain2007}
{Jain}, R. 2007, \apj, 656, 610, \dodoi{10.1086/510349}

\bibitem[{Jain \& Roberts(1991)}]{jain1991magnetoacoustic}
Jain, R., \& Roberts, B. 1991, Solar physics, 133, 263

\bibitem[{Korpi-Lagg {et~al.}(2022)Korpi-Lagg, Korpi-Lagg, Olspert, \&
  Truong}]{korpi2022}
Korpi-Lagg, M.~J., Korpi-Lagg, A., Olspert, N., \& Truong, H.-L. 2022, arXiv
  preprint arXiv:2205.04419

\bibitem[{Lee \& Roberts(1986)}]{lee1986behavior}
Lee, M., \& Roberts, B. 1986, The Astrophysical Journal, 301, 430

\bibitem[{Miles \& Roberts(1992)}]{miles1992magnetoacoustic}
Miles, A.~J., \& Roberts, B. 1992, Solar physics, 141, 205

\bibitem[{Miles(1960)}]{miles1960generation}
Miles, J.~W. 1960, Journal of Fluid Mechanics, 7, 469

\bibitem[{{Nye} \& {Thomas}(1976)}]{nye1976}
{Nye}, A.~H., \& {Thomas}, J.~H. 1976, \apj, 204, 573, \dodoi{10.1086/154205}

\bibitem[{{Pint{\'e}r}(2015)}]{pinter2015}
{Pint{\'e}r}, B. 2015, Journal of Astrophysics and Astronomy, 36, 15,
  \dodoi{10.1007/s12036-015-9331-3}

\bibitem[{Pint{\'e}r \& Erd{\'e}lyi(2011)}]{pinter2011}
Pint{\'e}r, B., \& Erd{\'e}lyi, R. 2011, Space science reviews, 158, 471

\bibitem[{Schunker {et~al.}(2005)Schunker, Braun, Cally, \&
  Lindsey}]{schunker2005}
Schunker, H., Braun, D.~C., Cally, P.~S., \& Lindsey, C. 2005, The
  Astrophysical Journal, 621, L149

\bibitem[{Singh {et~al.}(2015)Singh, Brandenburg, Chitre, \&
  Rheinhardt}]{singh2015pfmodes}
Singh, N.~K., Brandenburg, A., Chitre, S., \& Rheinhardt, M. 2015, Monthly
  Notices of the Royal Astronomical Society, 447, 3708

\bibitem[{Singh {et~al.}(2014)Singh, Brandenburg, \& Rheinhardt}]{singh2014}
Singh, N.~K., Brandenburg, A., \& Rheinhardt, M. 2014, The Astrophysical
  Journal Letters, 795, L8

\bibitem[{Singh {et~al.}(2016)Singh, Raichur, \& Brandenburg}]{singh2016high}
Singh, N.~K., Raichur, H., \& Brandenburg, A. 2016, The Astrophysical Journal,
  832, 120

\bibitem[{Singh {et~al.}(2020)Singh, Raichur, K{\"a}pyl{\"a}, Rheinhardt,
  Brandenburg, \& K{\"a}pyl{\"a}}]{singh2020}
Singh, N.~K., Raichur, H., K{\"a}pyl{\"a}, M.~J., {et~al.} 2020, Geophysical \&
  Astrophysical Fluid Dynamics, 114, 196

\bibitem[{Thomas(1983)}]{thomas1983magneto}
Thomas, J.~H. 1983, Annual Review of Fluid Mechanics, 15, 321

\bibitem[{Ulrich(1970)}]{ulrich1970five}
Ulrich, R.~K. 1970, The Astrophysical Journal, 162, 993

\bibitem[{{Yu}(1965)}]{yu1965}
{Yu}, C.~P. 1965, Physics of Fluids, 8, 650, \dodoi{10.1063/1.1761278}

\end{thebibliography}
